# Hyperelastic characterization via deep indentation

Mohammad Shojaeifard[1], Mattia Bacca[1*]

[1]Mechanical Engineering Department, Institute of Applied Mathematics
School of Biomedical Engineering
University of British Columbia, Vancouver BC V6T 1Z4, Canada

[*]Corresponding author: mbacca@mech.ubc.ca

## Abstract

Hyperelastic material characterization is crucial for understanding the behavior of soft materials—such as tissues, rubbers, hydrogels, and polymers—under quasi-static loading before failure. Traditional methods typically rely on uniaxial tensile tests, which require the cumbersome preparation of dumbbell-shaped samples for clamping in a uniaxial testing machine. In contrast, indentation-based methods, which can be conducted in-situ without sample preparation, have been underexplored. To characterize the hyperelastic behavior of soft materials, deep indentation is required, where the material response extends beyond linear elasticity. In this study, we perform finite element analysis to link the force ($F$) vs. indentation depth ($D$) curve with the hyperelastic behavior of a soft incompressible material, using a one-term Ogden model for simplicity. We identify three indentation regimes based on the ratio between indentation depth and the radius ($R$) of the spherical-tipped cylindrical indenter: (1) the Hertzian regime ($D < 0.1\,R$) with $F = ER^{0.5}D^{1.5}16/9$, (2) the parabolic regime ($D > 10\,R$) with $F = ED^2\beta$, where the indenter radius becomes irrelevant, and (3) an intermediate regime ($0.1\,R < D < 10\,R$) bridging the two extremes. We find that the Ogden strain-stiffening coefficient ($\alpha$) increases the parabolic indentation coefficient ($\beta$), allowing for the estimation of $\alpha$ from $\beta$. Furthermore, we observe that Coulomb friction increases $\beta$, potentially masking the effect of strain-stiffening for small $\alpha$. However, for $\alpha > 3$, friction has a negligible effect. Finally, our results show good agreement with experimental data, demonstrating that deep indentation can be an effective method for extracting hyperelastic properties from soft materials through *in-situ* testing.

## Introduction

Characterizing the hyperelastic response of soft materials is essential for a wide range of applications, from biomedical engineering to materials science and beyond. Traditional methods, such as uniaxial tensile tests, require destructive sample preparation, suffer from mechanical challenges related to clamping efficiency, and cannot be performed *in situ* or *in vivo*. Alternative methods, like parallel plate compression (McGarry, 2009; Okwara *et al.*, 2021; Mu *et al.*, 2025a), overcome the clamping issue but still require sample preparation, making them unsuitable for in situ applications. Indentation methods (Dagro *et al.*, 2019; He *et al.*, 2024; Li *et al.*, 2024) offer greater flexibility, as they can be performed *in situ* and *in vivo* and are non-destructive, eliminating the need for sample preparation. It is crucial to go beyond linear elastic regimes when characterizing hyperelastic materials to capture the full extent of their behavior. Current methods (Dagro *et al.*, 2019; Li *et al.*, 2024) typically focus on shallow indentation depths and require knowledge of the substrate's thickness and/or curvature. These geometric considerations activate the nonlinear behavior of the material but introduce complexity into the measurement. In this study, we present an alternative method based on deep indentation using small cylindrical probes with spherical tips. With this approach, the sampled substrate can be treated as a hyperelastic half-space due to the small size of the probe, effectively eliminating the variability introduced by substrate thickness and curvature. This method produces a straightforward force-depth response, divided into two distinct regimes: the *Hertzian (linear elastic) regime* at shallow depths and the *parabolic regime* at larger depths. The Hertzian regime is useful for characterizing the elastic modulus of the material, as it is independent of the nonlinear behavior, while the parabolic regime provides insight into additional elastic parameters. We describe the material using a 1-term Ogden incompressible model, which requires only two hyperelastic parameters: the elastic modulus $E$ and the strain stiffening coefficient $\alpha$. By conducting finite element analysis (FEA), we establish correlations between the Hertzian behavior and $E$, and between the parabolic behavior and $\alpha$. The parabolic regime is also influenced by the friction coefficient, $f$, which has important implications in both indentation and puncture mechanics (Fregonese *et al.*, 2022). Our results show that both $\alpha$ and $f$ increase the parabolic force response, with friction sometimes overshadowing strain stiffening. Fortunately, the effect of friction becomes negligible for $\alpha > 3$.

We validate this method through a series of uniaxial tension and deep indentation experiments on four soft materials: Ecoflex® 00-10, Ecoflex® 00-30, Mold Star™ 16 Fast, and porcine skin. The hyperelastic parameters extracted from both methods show good agreement, supporting the robustness of the model and the proposed fitting strategy. A key observation is the ambiguity around the sign of $\alpha$, as materials like brain tissue (Budday *et al.*, 2020) exhibit negative values for $\alpha$. However, this ambiguity is less problematic in deep indentation, thus highlighting the robustness of our proposed method. We also discuss the limitations of using a small set of hyperelastic parameters, which restricts our ability to capture volumetric compressibility—an important factor in indentation and cutting of soft materials (Fregonese *et al.*, 2023; Goda *et al.*, 2025).

## Uniaxial Tension

We describe the material's elastic response via a 1-term Ogden hyperelastic incompressible model (Ogden, 1972) for this study, where the strain energy density (SED) is

$$\psi = \frac{2E}{3\alpha^2}\left(\bar{\lambda}_1^\alpha + \bar{\lambda}_2^\alpha + \bar{\lambda}_3^\alpha - 3\right) \tag{1}$$

Here, $E$ is the (zero-strain) Young's modulus, $\alpha$ is the strain-stiffening coefficient (Ogden, 1972), and $\bar{\lambda}_i = \lambda_i J^{-1/3}$ represents the deviatoric component of the principal stretch $\lambda_i$, with $J$ the swelling ratio. Note that $\lambda_i = dl_i/dL_i$, where $dl_i$ and $dL_i$ are the current and reference (unloaded) lengths of a unit segment in the principal direction $i$, and $J = \lambda_1\lambda_2\lambda_3 = dv/dV$ is the volume swelling ratio, where $dv$ and $dV$ are the current and reference unit volumes.

Considering uniaxial tension, we have that $\lambda_1 = \lambda = l/L$, with $l$ and $L$ the sample length in the loaded and unloaded states in the pulling direction. Here, incompressibility gives us $J = \lambda_1\lambda_2\lambda_3 = 1$, so that $\lambda_2 = \lambda_3 = \lambda^{-1/2}$. Also $\bar{\lambda}_i = \lambda_i$ for all three principal directions. The SED becomes then

$$\psi = \frac{2E}{3\alpha^2}\left(\lambda^\alpha + 2\lambda^{-\frac{\alpha}{2}} - 3\right) \tag{2}$$

The nominal (engineering) puling stress $S = F/A_0$, with $F$ puling force and $A_0$ initial cross-section area, is $S = \partial\psi/\partial\lambda$, giving

$$S = \frac{2E}{3\alpha}\left(\lambda^{\alpha-1} - \lambda^{-\frac{\alpha}{2}-1}\right) \tag{3}$$

Figures 1-*left* and 2 presents the stress-stretch plots from Eq. (3) with log axes. At large stretch, Eq. (3) gives the asymptotic law

$$S \approx \frac{2E}{3\alpha}\lambda^{\alpha-1} \tag{4}$$

for positive $\alpha$, and

$$S \approx \frac{2E}{3|\alpha|}\lambda^{\frac{|\alpha|}{2}-1} \tag{5}$$

for negative $\alpha$. Albeit less common than the case of positive $\alpha$, negative $\alpha$ is found in the characterization of brain matter (Budday *et al.*, 2020).

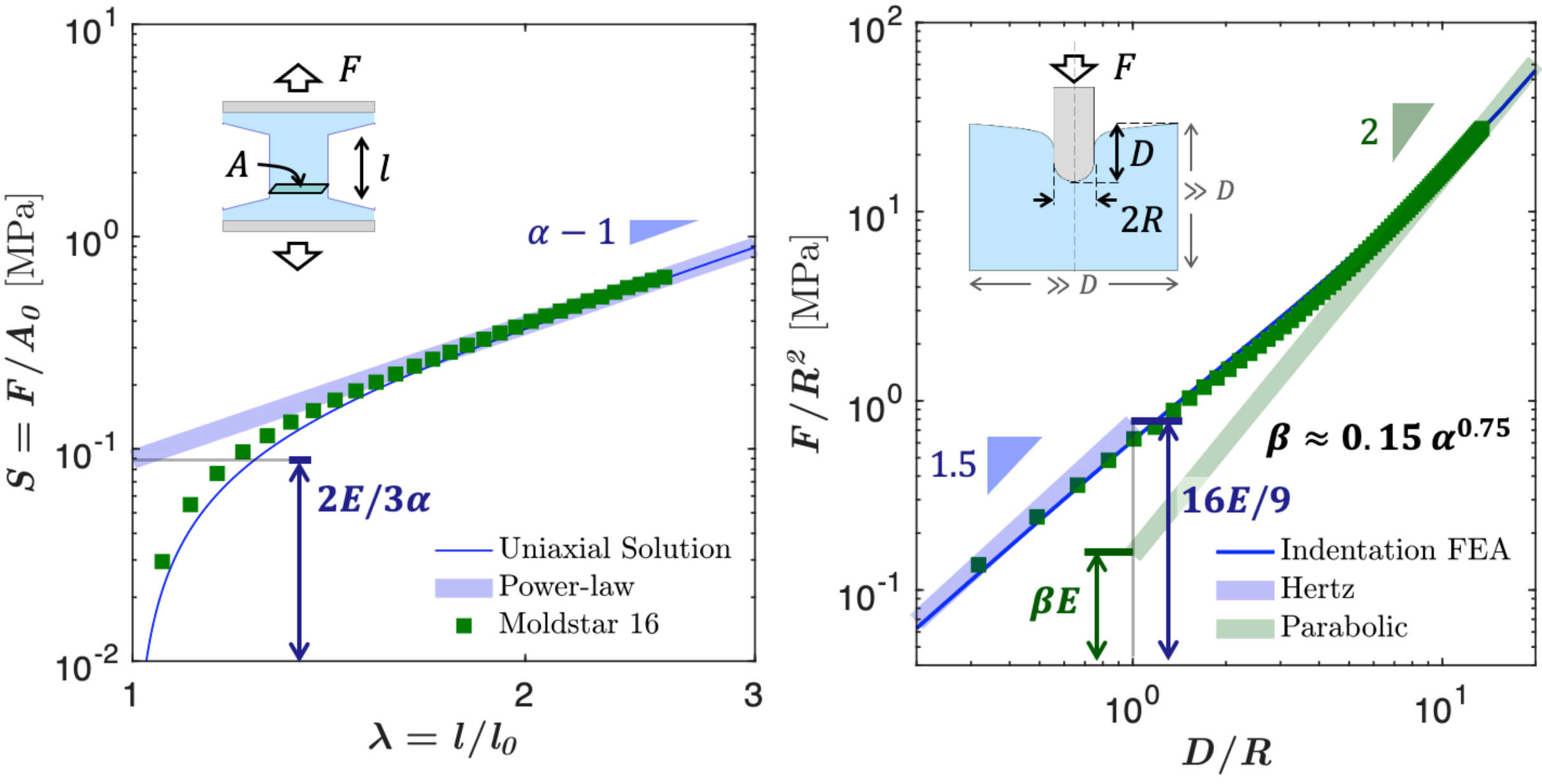

**Figure 1**: Schematic comparison between traditional uniaxial tension (left) and the deep indentation method (right) proposed in this study for hyperelastic characterization using spherically-tipped rigid cylinders. Plots show experimental results on Mold Star 16 Fast silicone (Moldstar 16: green squares), compared with theoretical predictions from the closed-form uniaxial solution (left) and finite element analysis (FEA) for indentation (right) (solid lines). Both plots use logarithmic axes in *x* and *y*, highlighting power-law behavior as linear trends in log-log scale (faint thick lines). In both cases, the slopes and intercepts of the trend lines are used to extract the hyperelastic parameters: the elastic modulus $E$ and the Ogden strain-stiffening coefficient $\alpha$. In uniaxial tension, a single power-law emerges at large stretch $\lambda$, with slope $\alpha - 1$ and intercept $2E/3\alpha$ at $\lambda = 1$. The *y*-axis reports engineering stress $S$. In deep indentation, two regimes appear: at shallow depth-to-radius ratio $D/R$, the response follows Hertzian

mechanics with slope 1.5 and intercept $16E/9$ at $D = R$; at greater depths, a parabolic regime emerges with slope 2 and intercept $\beta E$ at $D = R$, where $\beta$ correlates with $\alpha$ via the empirical relation $\beta \approx 0.15\ \alpha^{0.75}$. Note that in this regime, we have $F \sim ED^2$, where the dependency on $R$ is lost.

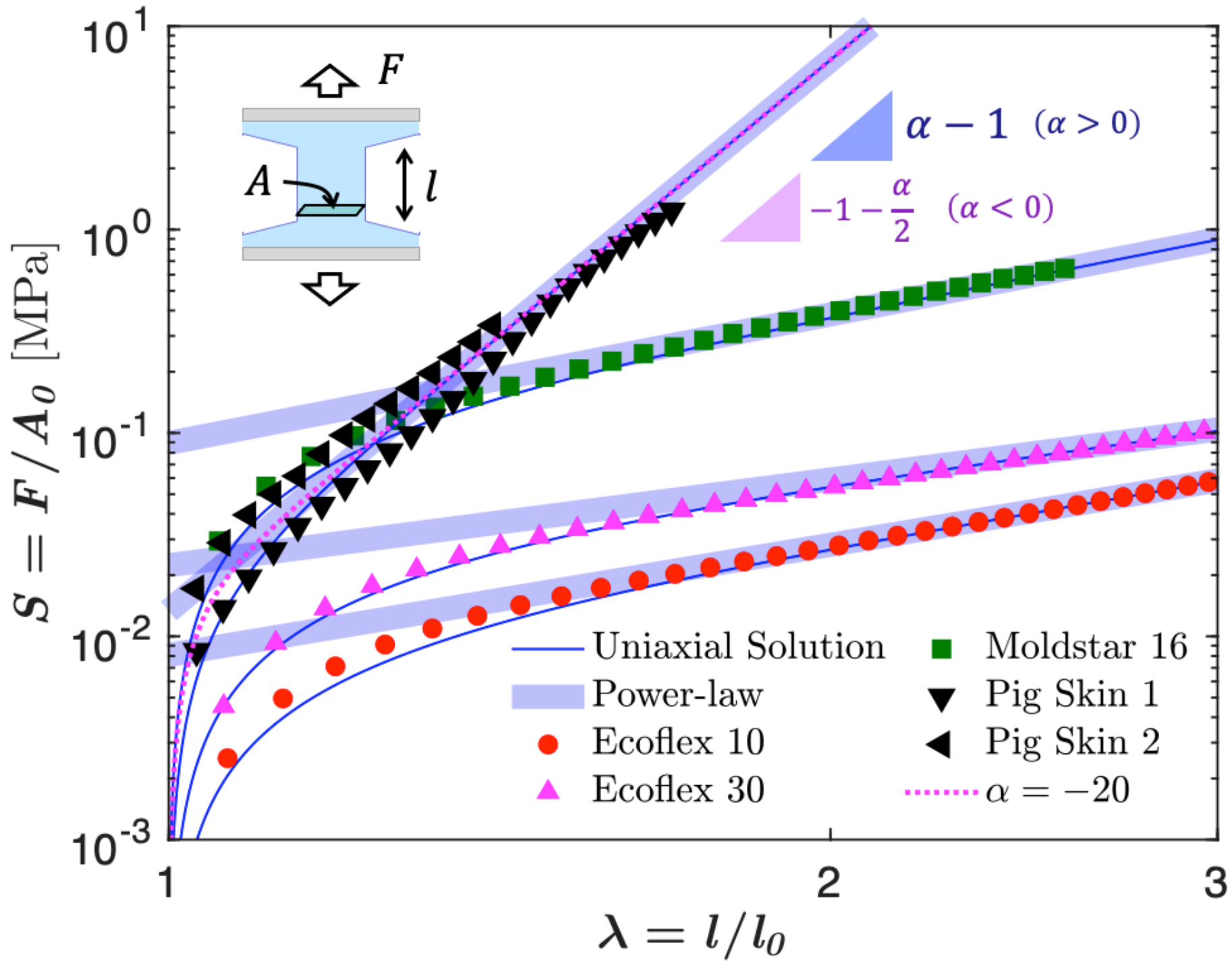

**Figure 2**: Uniaxial tests comparing experimental results with the closed-form solution from Ogden's hyperelastic model (Eq. (3), blue lines) and the power-law trend (Eq. (4), faint thick lines). The tested materials are Ecoflex 10 and 30, Mold Star 16 Fast (Smooth-On), and pig skin. For the latter, "1" and "2" denote two sets of samples tested along orthogonal directions to highlight the material's anisotropy. All datasets represent averages over three tests with different samples. Parameter extraction follows the procedure illustrated in Figure 1 and its caption. Note that the slope of the trend line is $\alpha - 1$ only when $\alpha$ is positive; for negative $\alpha$, the slope becomes $-1 - \alpha/2$ (see Eqs. (4) and (5)). The intercept is $2E/3|\alpha|$ in both cases. The ambiguity in the sign of $\alpha$, where both $\alpha = 10$ and $\alpha = -20$ yield the same slope of 9, is illustrated with the magenta dashed line, which corresponds to $\alpha = -20$ and a modulus $2E$. The same slope would result from $\alpha = 10$ with modulus $E$, which is the correct choice in this case, as pig skin is known to have a positive $\alpha$.

In Figures 1-*left* and 2 we compare our predictions of Eq. (3), as well as Eqs (4) and (5), with experiments on four materials: Ecoflex 10 and 30, Mold Star 16 Fast silicone elastomers (Smooth-On), and pig skin. The parameter extraction procedure, sketched in Figure 1-*left* and detailed below, relies on fitting the experimental data at large stretches using Eq. (4), appropriate for materials with positive $\alpha$. From the slope of the power-law trend ($\alpha - 1$), we extract the Ogden parameter $\alpha$, and from the intercept at $\lambda = 1$ ($2E/3\alpha$), we determine the modulus $E$. The extracted parameters are summarized in Table 1 and are consistent with previous measurements reported in the literature (Jansen et al., 1958; Shergold et al., 2005–2006; Joodaki et al., 2018; Liao et al., 2020; Marechal et al., 2021).

In Figure 2, we compare Eqs. (4) and (5) in fitting the response of pig skin, illustrating the ambiguity in determining the sign of $\alpha$. A slope of 9 yields $\alpha = 10$ under Eq. (4) (blue line), whereas Eq. (5) gives $\alpha = -20$ (dashed magenta line).

While Eq. (4) leads to a modulus of $E = 0.28\ MPa$, as reported in Table 1, Eq. (5) would imply a modulus nearly twice as large. A practical way to resolve this ambiguity is to compare the predicted modulus with an estimate of $E$ at infinitesimal strains, obtained by fitting the initial portion of the stress–strain curve with Hooke's law, though this can be limited by load cell sensitivity. Another approach is to perform additional tests under different loading conditions. As will be shown, deep indentation offers a promising alternative. Parallel-plate compression could also be used, although it often suffers from frictional effects at the sample–plate interface.

**Table 1**: Estimated hyperelastic parameters, modulus $E$ and Ogden parameter $\alpha$, for the four materials tested, using both uniaxial tension and deep indentation. Parameter extraction follows the procedure illustrated in Figure 1. The relative errors $e_E$ and $e_\alpha$ quantify the discrepancy between the two methods.

| | ***Uniaxial Tension*** | | ***Deep Indentation*** | | | ***Error*** | |
|---|---|---|---|---|---|---|---|
| | $E\ [MPa]$ | $\alpha$ | $E\ [MPa]$ | $\beta$ | $\alpha$ | $e_E$ | $e_\alpha$ |
| **Ecoflex 10** | 0.034 | 2.8 | 0.038 | 0.34 | 3.04 | 11% | 8.4% |
| **Ecoflex 30** | 0.081 | 2.4 | 0.080 | 0.31 | 2.64 | 0.1% | 10.1% |
| **Moldstar 16** | 0.42 | 3.1 | 0.43 | 0.36 | 3.19 | 3% | 2.8% |
| **Pig Skin** | 0.28 | 10 | 0.22 | 0.83 | 9.81 | 22.8% | 1.9% |

*Experimental Method (Uniaxial Tension)*

Uniaxial tension tests were performed on dogbone-shaped specimens (25 mm gauge length × 15 mm width × 2 mm thickness, so $A_0 = 30\ mm^2$) under quasi-static loading at a nominal strain rate of $\dot{\varepsilon} = 4 \cdot 10^{-3} s^{-1}$, at controlled temperature (22 ± 1 °C) and relative humidity (45 ± 5%). Each material, Ecoflex® 00-10, Ecoflex® 00-30, Mold Star™ 16 Fast (Smooth-On, Inc.), and fresh porcine skin, was tested using three replicates ($n = 3$). For porcine skin, two sets of three samples were cut in orthogonal directions (denoted as Pig Skin 1 and Pig Skin 2) to reveal the material's anisotropy.

Force was measured using the load cell of an Instron® universal testing machine, while stretch $\lambda$ was obtained optically by tracking surface deformation from 4K-resolution video recordings (100 Hz). Horizontal fiducial lines drawn on the specimen surface were tracked frame-by-frame (SI: Figure S2) using *Tracker* video analysis software (Open-Source Physics, developed by Doug Brown), allowing local displacements to be measured in the gauge region. Engineering stress was calculated from the force and initial cross-sectional area. Reported stress–stretch curves represent the mean response for each set, with sample-to-sample variation below 3%.

**Deep Indentation via Spherically-tipped Cylinders**

Let us now analyze deep indentation using a spherically-tipped rigid cylinder. The key variables are the indenter radius $R$, indentation depth $D$, the material's elastic modulus $E$, the Ogden strain stiffening coefficient $\alpha$ (see Eq. (1)-(5)), and the Coulomb friction coefficient $f = \tau_f/p$, where $\tau_f$ is the frictional shear stress and $p$ is the contact pressure, both at the indenter-specimen interface.

The force-depth response of a single-term incompressible Ogden material was simulated using finite-element analysis (FEA) in Abaqus/Explicit 2024 (dynamic, explicit solver). A 2D axisymmetric model was used, with a rigid spherical indenter pressing into a cylindrical elastomer sample of radius $B$ and height $H$ equal to $100R$ (SI: Figure S1). Indentation was imposed via a prescribed vertical displacement applied to a reference point coupled to the indenter, reaching a maximum depth of $D/R = 25$, unless numerical instability prevented such depth ($\alpha = 2$) or the parabolic trend was reached at shallower depths. Frictionless and frictional contact conditions were applied with Coulomb friction coefficients $f = 0, 0.1, 1$. Simulations were conducted for various strain-stiffening parameters $\alpha = 2,\ 3,\ 5,\ 9,\ -9,\ -20$. The indentation speed was selected to maintain quasi-static conditions, keeping the kinetic-to-internal energy ratio below 2%. The reaction force was extracted and plotted versus indentation depth (Figure 3).

At small indentation depths ($D \ll R$), the indentation force $F$ correlates with $D$ as follows, based on Hertzian theory

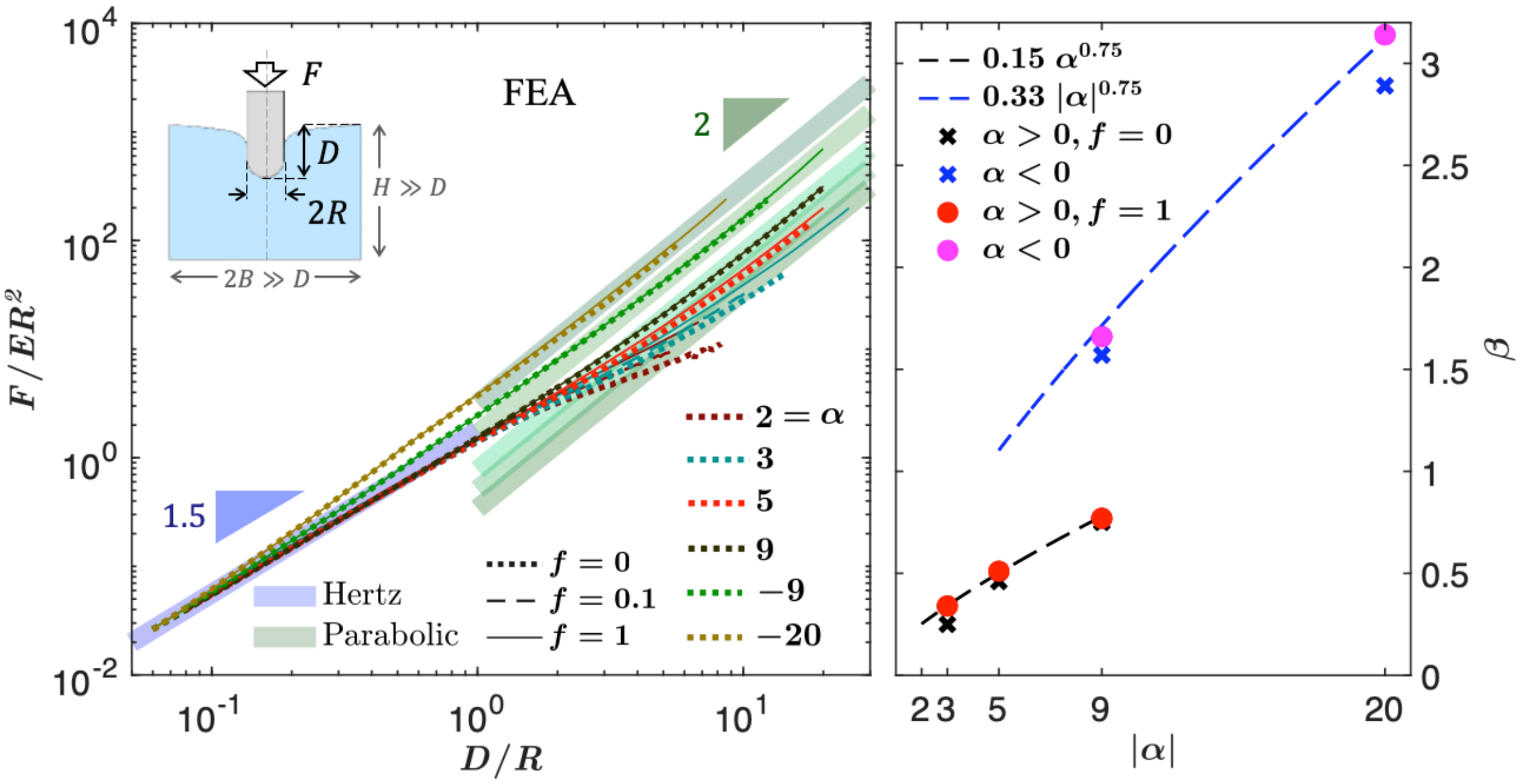

**Figure 3**: Results from finite element analysis (FEA) using the Abaqus/Explicit module. *Left*: Simulated indentation curves for various Ogden strain-stiffening parameters, $\alpha$, showing the Hertz and parabolic regimes. *Right*: Correlation between the parabolic coefficient $\beta$ (from Eq. (7)) and the Ogden parameter $\alpha$. The empirical trends $\beta \approx 0.15\alpha^{0.75}$ for $\alpha > 0$, and $\beta \approx 0.33|\alpha|^{0.75}$ for $\alpha < 0$, are derived from FEA data. While in uniaxial tension a negative $\alpha$ corresponds to a stiffer response (requiring twice the modulus to match the slope), under indentation it leads to a softer response (requiring half the modulus). Simulations include Coulomb friction coefficients $f = 0$, 0.1, and 1; friction effects are negligible for large $\alpha$.

$$\frac{F}{ER^2} \approx \frac{16}{9}\left(\frac{D}{R}\right)^{\frac{3}{2}} \tag{6}$$

This relationship generally holds for $D < 0.1\,R$, as shown by both our FEA results (Figures 1-*right* and 3-*left*) and experiments (Figure 4). Notably, the force-depth response is independent of friction and strain stiffening, as all curves align with the Hertzian log-log line for any values of $f$ and $\alpha$.

At larger depths ($D \gg R$, assumed for $D > 10\,R$), both FEA and experiments indicate that $F \sim ED^2$. The dependence on $R$ vanishes as the material's resistance near the contact region depends only on the indentation depth $D$, not on the radius $R$. The relationship is then

$$\frac{F}{ER^2} \approx \beta\left(\frac{D}{R}\right)^2 \tag{7}$$

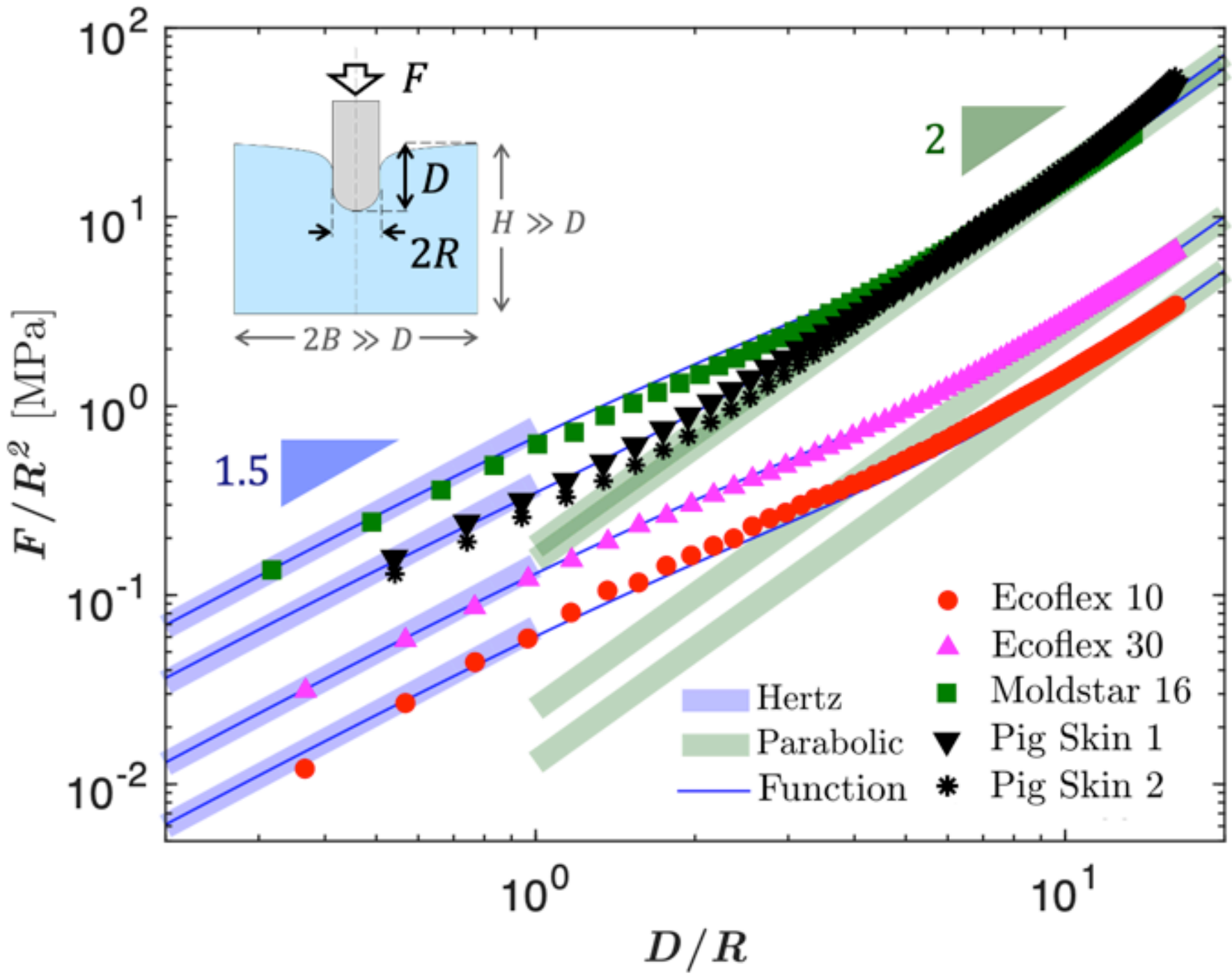

**Figure 4**: Results of indentation tests showing force normalized by radius squared, $F/R^2$, versus normalized depth, $D/R$, plotted in log-log scale for spherically tipped cylindrical indenters. Experimental data are compared with theoretical predictions, including the Hertz and Parabolic fits (faint thick lines) described in Figure 1 and its caption, as well as the empirical function

given in Eq. (9) (blue lines). Tests were conducted on the same materials as in Figure 2, using two samples per material and reporting the average response, except for pig skin, for which individual datasets are shown to illustrate the inherent variability typical of biological materials.

where $\beta$ depends on the material's hyperelastic behavior, via $\alpha$ and $E$, and frictional interactions, via $f$. For $\alpha = 2$, corresponding to neo-Hookean, the indentation response does not exhibit a parabolic behavior within the explored depth range (Figure 3). While it is possible that at significantly larger indentation depths, the behavior for $\alpha = 2$ aligns with Eq. (7), the critical depth required for this transition might be indefinitely large. The limitations in the maximum explorable depth $D$ via FEA are due to element distortion, a numerical issue that is particularly exacerbated in the case of $\alpha = 2$, likely due to the strain-softening behavior of neo-Hookean materials. This strain-softening leads to large deformations that challenge the numerical stability of the model.

Note that, for positive values of α within the explored range, we find $\beta < 1$, whereas $16/9 = 1.78$. This indicates that the indentation resistance beyond the Hertzian regime, while tending toward the parabolic regime, decreases rather than increases. Such a reduction in resistance is consistent with recent findings by Mu et al. (2025).

Friction, which only affects the parabolic regime described in Eq. (7), appears to produce negligible effects for larger values of $\alpha$, as the curves for $f = 0$ (solid lines) and $f = 1$ (dashed lines) remain closely aligned.

Figure 3-*right* shows the correlation between $\alpha$ and $\beta$, fitted as:

$$\beta \approx 0.15\ \alpha^{0.75}, \text{ for } \alpha > 0 \quad (8a)$$

$$\beta \approx 0.33\ |\alpha|^{0.75}, \text{ for } \alpha < 0 \quad (8b)$$

Negative $\alpha$ results in a roughly 2-fold larger $\beta$ compared to positive $\alpha$, as negative $\alpha$ increases strain stiffening under compression, which dominates during indentation. Larger $\beta$ values also cause a transition from Hertz to parabolic behavior at smaller $D/R$, suggesting that brain matter may exhibit parabolic behavior at relatively shallow depths (Budday *et al.*, 2020).

It is important to note that extracting $\alpha$ from deep indentation tests alone may not allow one to distinguish between positive and negative values of $\alpha$, since a given value of $\beta$ can correspond to two distinct $\alpha$ values, one positive and one negative.

This sign ambiguity can compromise the accurate identification of $\alpha$. For example, while in uniaxial tests $\alpha = -20$ may be confused with $\alpha = 10$ (see Figure 2), in deep indentation, equating Eqs. (8a) and (8b) reveals that $\alpha = -3.49$ produces the same $\beta$ as $\alpha = 10$. Fortunately, this ambiguity is resolved by combining uniaxial tension with deep indentation, enabling a unique and accurate estimation of both $\alpha$ and $E$.

Notably, in Figure 3-*right*, we observe that while the effect of friction diminishes with increasing $\alpha$ for $\alpha > 0$, the opposite trend occurs for $\alpha < 0$, where friction becomes more influential.

To capture the indentation response across both shallow and deep regimes, we combine Eqs. (6) and (7) into the following empirical expression

$$\frac{F}{R^2} = \frac{16}{9} E \left(\frac{D}{R}\right)^{\frac{3}{2}} \exp\left(-\beta \frac{D}{R}\right) + \beta E \left(\frac{D}{R}\right)^2 \quad (9)$$

This equation, shown as a solid blue line in Figure 4, blends the Hertzian regime with an exponential decay and the parabolic regime. Figure 4 demonstrates how Eq. (9) closely matches our experimental measurements, which are described in the next section. For reference, the individual Hertzian and parabolic trends from Eqs. (6) and (7) are also plotted as faint thick lines, showing good agreement with the experimental data.

*Experimental Method (Deep Indentation)*

Deep indentation tests were performed on cylindrical specimens (radius $B = 40\ mm$, height $H = 45\ mm$, Figure 4) of Ecoflex® 00-10, Ecoflex® 00-30, Mold Star™ 16 Fast (Smooth-On, Inc.), and fresh porcine skin, the same materials tested in uniaxial tension. Elastomer samples were cast into 3D-printed molds to the required dimensions. Porcine skin specimens were prepared by trimming subcutaneous fat to a thickness of $\sim 2\ mm$, cutting circular disks (radius $B = 40\ mm$), and stacking 14 layers bonded with cyanoacrylate to reach a height of $H = 45\ mm$.

All tests were conducted on an Instron® universal testing machine at a constant crosshead speed of $0.1\ mm/s$. A rigid stainless-steel indenter with a spherical tip radius of $R = 0.5\ mm$ was centered on the sample surface and advanced to a depth of $6.8 - 8.2\ mm$ (*i.e.*, $D/R = 13.6$, Moldstar 16, and $= 16.3$ all others), then retracted at the same rate. The indenter consisted of a long steel shaft and a steel ball of equal diameter glued to its end (SI: Figure S2). Force–displacement data were recorded at 5 Hz.

For each elastomer, two tests were performed, and the resulting force–displacement curves were averaged; the deviation between runs was within 4.5%. For porcine skin, both curves are reported without averaging to reflect biological variability.

## **Discussion and Conclusion**s

Hyperelastic characterization of soft materials via deep indentation offers both significant opportunities and practical challenges. The ability to extract mechanical parameters in situ using simple indentation protocols provides an attractive and accessible alternative to traditional bulk testing methods. In this study, we have demonstrated the feasibility of this approach across a broad range of soft materials using a simplified model based on just two parameters: the elastic modulus $E$ and the Ogden strain-stiffening coefficient $\alpha$.

Our method focuses on capturing the indentation response across two distinct regimes: the Hertzian regime at shallow depths and the parabolic regime at larger indentations. These two power-law behaviors are not specific to our constitutive choice, but rather reflect universal mechanical principles: the Hertzian regime arises from linear elasticity at small strains, while the parabolic regime emerges from geometric considerations when the indentation depth greatly exceeds the indenter radius. What enables parameter identification in our approach is the simplicity of the one-term Ogden model, which allows the elastic modulus $E$ and strain-stiffening parameter $\alpha$ to be mapped directly onto these two regimes. For more complex hyperelastic models, such as compressible formulations with a finite bulk modulus $K$, multi-term Ogden models (Ogden, 1972), or the Arruda–Boyce model (1993), the power-law behavior in these regimes still exists, but the number of parameters exceeds the number of independent trends that can be reliably extracted from log–log data alone. In such cases, additional experimental strategies may be required to improve parameter identifiability, for example, varying probe geometry or using multiprobe configurations, such as microneedle arrays designed to probe coupling effects arising from the proximity of multiple needles (Jahan et al., 2025).

To validate our model, we conducted uniaxial and deep indentation experiments on four representative soft materials: Ecoflex® 00-10, Ecoflex® 00-30, Mold Star™ 16 Fast, and porcine skin. For each material, the hyperelastic parameters $E$ and $\alpha$ were independently extracted from both testing methods and compared. We found close agreement between the two approaches, with a maximum discrepancy of 11% for the three elastomers. For porcine skin, which is anisotropic, the maximum error increased to 22%, likely due to direction-dependent mechanical behavior that cannot be fully captured by isotropic models. These results provide strong experimental support for the proposed indentation-based method and confirm its ability to reliably estimate hyperelastic parameters using minimal sample preparation and compact test setups.

A key challenge in this method is ensuring sufficient indentation depth to activate the parabolic regime without approaching the puncture threshold. Fregonese et al. (2021) showed that the critical puncture ratio $D_c/R$ can exceed 100 in soft gels, based on analysis of data from Fakhouri et al. (2015) and Rattan et al. (2018). In our experiments, the parabolic trend became apparent at $D/R \approx 6$ for Mold Star™ 16 Fast and porcine skin, materials with higher $\alpha$, and required $D/R \gtrsim 10$ for Ecoflex® 00-10 and 00-30. Since the critical $D_c/R$ increases with decreasing indenter radius, using smaller indenters increases the likelihood of capturing the parabolic regime before puncture occurs. This strategy is particularly useful in in-situ or in-vivo applications, where the sample dimensions $B$ and $H$ may be unknown or constrained, and the elastic half-space assumption may be violated. By reducing the indenter radius $R$, the critical depth and corresponding minimum required sample dimensions are also reduced.

Another key aspect is the presence of a "skin" layer, typical of biological tissues (Zhang et al., 2024) and sometimes present in synthetic materials due to surface oxidation or environmental exposure. This skin may affect both the Hertzian and parabolic responses, meaning deep indentation may reflect surface rather than bulk properties. However, if the skin is thin compared to the indentation depth, its influence is likely minimized, allowing access to more representative bulk behavior.

Frictional effects can also influence the accuracy of parameter extraction. As shown in our FEA results, friction primarily affects the parabolic regime for strain-softening materials (i.e., those with low $\alpha$), while materials with higher $\alpha$ show minimal sensitivity to friction. This observation is consistent with the idea that materials with intrinsic strain stiffening are better suited for reliable indentation-based characterization.

Finally, achieving high-resolution force measurements is essential. In our experiments, minimum forces fell within the milli-Newton range, requiring sensitive load cells to resolve the initial Hertzian regime. These practical considerations—including probe size and force resolution—must be carefully addressed to ensure reproducible and meaningful results.

In summary, this study validates deep indentation as a compact, minimally invasive technique to extract hyperelastic properties in soft materials. Theoretical predictions based on finite-element simulations were shown to match experimental data, and material parameters derived from indentation showed strong agreement with those from standard uniaxial testing. While limited in scope by model simplicity and assumptions (e.g., isotropy and incompressibility), this method holds strong promise for rapid mechanical characterization, especially in biomedical or field-deployable contexts.

## Acknowledgments

We gratefully acknowledge insightful discussions with Norman Fleck, Yiting Wu, as well as Zhenwei Ma, who also provided the experimental equipment. This work was supported by the Natural Sciences and Engineering Research Council of Canada (NSERC), grant RGPIN-2025-07085.

## References

Arruda, E.M. and Boyce, M.C., 1993. A three-dimensional constitutive model for the large stretch behavior of rubber elastic materials. *Journal of the Mechanics and Physics of Solids*, *41*(2), pp.389-412.

Budday, S., Ovaert, T.C., Holzapfel, G.A., Steinmann, P. and Kuhl, E., 2020. Fifty shades of brain: a review on the mechanical testing and modeling of brain tissue. *Archives of Computational Methods in Engineering*, *27*, pp.1187-1230.

Dagro, A.M. and Ramesh, K.T., 2019. Nonlinear contact mechanics for the indentation of hyperelastic cylindrical bodies. *Mechanics of Soft Materials*, *1*(1), p.7.

Fakhouri, S., Hutchens, S.B. and Crosby, A.J., 2015. Puncture mechanics of soft solids. *Soft matter*, *11*(23), pp.4723-4730.

Fischer-Cripps, A.C., 2007. *Introduction to contact mechanics* (Vol. 101). New York: Springer.

Fregonese, S. and Bacca, M., 2021. Piercing soft solids: a mechanical theory for needle insertion. *Journal of the Mechanics and Physics of Solids*, *154*, p.104497.

Fregonese, S. and Bacca, M., 2022. How friction and adhesion affect the mechanics of deep penetration in soft solids. *Soft Matter*, *18*(36), pp.6882-6887.

Fregonese, S., Tong, Z., Wang, S. and Bacca, M., 2023. Theoretical puncture mechanics of soft compressible solids. *Journal of Applied Mechanics*, *90*(11), p.111003.

Goda, B.A. and Bacca, M., 2025. Cutting Mechanics of Soft Compressible Solids–Force-radius scaling versus bulk modulus. *Mechanics of Materials*, p.105271.

He, D., Malu, D. and Hu, Y., 2024. A comprehensive review of indentation of gels and soft biological materials. *Applied Mechanics Reviews*, *76*(5), p.050802.

Hertz, H., 1882. On the contact of rigid elastic solids and on hardness, chapter 6: Assorted papers by H. Hertz. *MacMillan, New York*.

Jahan, S., Jain, A., Fregonese, S., Hu, C., Bacca, M. and Panat, R., 2025. Bed-of-Nails Effect: Unraveling the Insertion Behavior of Aerosol Jet 3D Printed Microneedle Array in Soft Tissue. *Extreme Mechanics Letters*, p.102301.

Jansen, L.H. and Rottier, P.B., 1958. Some mechanical properties of human abdominal skin measured on excised strips: a study of their dependence on age and how they are influenced by the presence of striae. *Dermatology*, *117*(2), pp.65-83.

Johnson, K.L., 1987. *Contact mechanics*. Cambridge university press.

Joodaki, H. and Panzer, M.B., 2018. Skin mechanical properties and modeling: A review. *Proceedings of the Institution of Mechanical Engineers, Part H: Journal of Engineering in Medicine*, *232*(4), pp.323-343.

Li, L. and Masen, M., 2024. A new method for determining the ogden parameters of soft materials using indentation experiments. *Journal of the mechanical behavior of biomedical materials*, *155*, p.106574.

Liao, Z., Hossain, M. and Yao, X., 2020. Ecoflex polymer of different Shore hardnesses: Experimental investigations and constitutive modelling. *Mechanics of Materials*, *144*, p.103366.

Long, R., Hui, C.Y., Gong, J.P. and Bouchbinder, E., 2021. The fracture of highly deformable soft materials: A tale of two length scales. *Annual Review of Condensed Matter Physics*, *12*(1), pp.71-94.

Marechal, L., Balland, P., Lindenroth, L., Petrou, F., Kontovounisios, C. and Bello, F., 2021. Toward a common framework and database of materials for soft robotics. *Soft robotics*, *8*(3), pp.284-297.

McGarry, J.P., 2009. Characterization of cell mechanical properties by computational modeling of parallel plate compression. *Annals of biomedical engineering*, *37*, pp.2317-2325.

Mu, T., Li, R., Linghu, C., Liu, Y., Leng, J., Gao, H. and Hsia, K.J. 2025. Nonlinear Contact Mechanics of Soft Elastic Spheres Under Extreme Compression. *Available at SSRN 5164995*.

Mu, T., Linghu, C., Liu, Y., Leng, J., Gao, H. and Hsia, K.J., 2025. Universal Scaling Laws for Deep Indentation Beyond the Hertzian Regime. *arXiv preprint arXiv:2506.11461*.

Ogden, R.W., 1972. Large deformation isotropic elasticity–on the correlation of theory and experiment for incompressible rubberlike solids. *Proceedings of the Royal Society of London. A. Mathematical and Physical Sciences*, *326*(1567), pp.565-584.

Okwara, C.K., Vaez Ghaemi, R., Yu, C., Le, M., Yadav, V.G. and Frostad, J.M., 2021. The mechanical properties of neurospheres. *Advanced Engineering Materials*, *23*(8), p.2100172.

Rattan, S. and Crosby, A.J., 2019. Effect of polymer volume fraction on fracture initiation in soft gels at small length scales. *ACS Macro Letters*, *8*(5), pp.492-498.

Shergold, O.A. and Fleck, N.A., 2005. Experimental investigation into the deep penetration of soft solids by sharp and blunt punches, with application to the piercing of skin.

Shergold, O.A., Fleck, N.A. and Radford, D., 2006. The uniaxial stress versus strain response of pig skin and silicone rubber at low and high strain rates. *International journal of impact engineering*, *32*(9), pp.1384-1402.

Zhang, B., Baskota, B. and Anderson, P.S., 2024. Being thin-skinned can still reduce damage from dynamic puncture. *Journal of the Royal Society Interface*, *21*(219), p.20240311.

# SI - Hyperelastic characterization via deep indentation

Mohammad Shojaeifard[1], Mattia Bacca[1*]
[1]Mechanical Engineering Department, Institute of Applied Mathematics
School of Biomedical Engineering
University of British Columbia, Vancouver BC V6T 1Z4, Canada
[*]Corresponding author: mbacca@mech.ubc.ca

**Finite Element Analysis**

Finite-element simulations were carried out in Abaqus/Explicit 2024 using a two-dimensional axisymmetric model of a rigid indenter pressing into a cylindrical elastomer sample. The sample, whose radius $B$ and height $H$ were both set to $100R$ (with $R$ the indenter radius), was meshed with 117,907 CAX4R elements (4-node bilinear axisymmetric quadrilaterals with reduced integration and hourglass control), while the indenter was represented by 166 RAX2 rigid link elements, for a total of 118,073 elements. Symmetry was enforced along the axisymmetric (left) edge ($u_x = 0$); the right edge was prevented from moving radially ($u_x = 0$), and the bottom face was fixed in the vertical direction ($u_y = 0$). The indenter geometry was coupled to a single reference point (RP), to which a prescribed vertical displacement was applied. A schematic of the geometry, boundary conditions, and mesh grading is shown, with a fine mesh in the contact region that becomes progressively coarser toward the boundaries. The sample material was modeled as a single-term incompressible Ogden hyperelastic solid. A hard contact condition was used at the indenter–sample interface, and friction was modeled using a Coulomb contact law, with simulations performed for friction coefficients $f = 0,\ 0.1,\ 1$. This parametric variation allowed us to assess the influence of friction on the force–displacement response and isolate its effect from that of material nonlinearity. A constant, low indentation velocity was assigned to the RP so that a total penetration depth of $25R$ was reached. The speed was chosen to maintain a kinetic-to-internal energy ratio below 2%**,** ensuring a quasi-static response. Mesh convergence was verified by halving the element size in the contact region until the peak force at $D/R = 25$ changed by less than 2%. The reaction force at the RP was recorded and plotted versus indentation depth.

**Uniaxial and Indentation Experiments**

Figure S2 shows photographs of the specimens used in the experiments, specifically Ecoflex® 00-30. On the left, the Instron® clamps are shown pulling the dogbone-shaped sample, which is marked with horizontal lines tracked optically to accurately measure stretch. On the right, the indenter used for the indentation tests is shown. It has a tip and shaft radius of $R = 0.5\ mm$, consisting of a long steel cylinder with a steel ball glued to its end, both components having the same diameter.

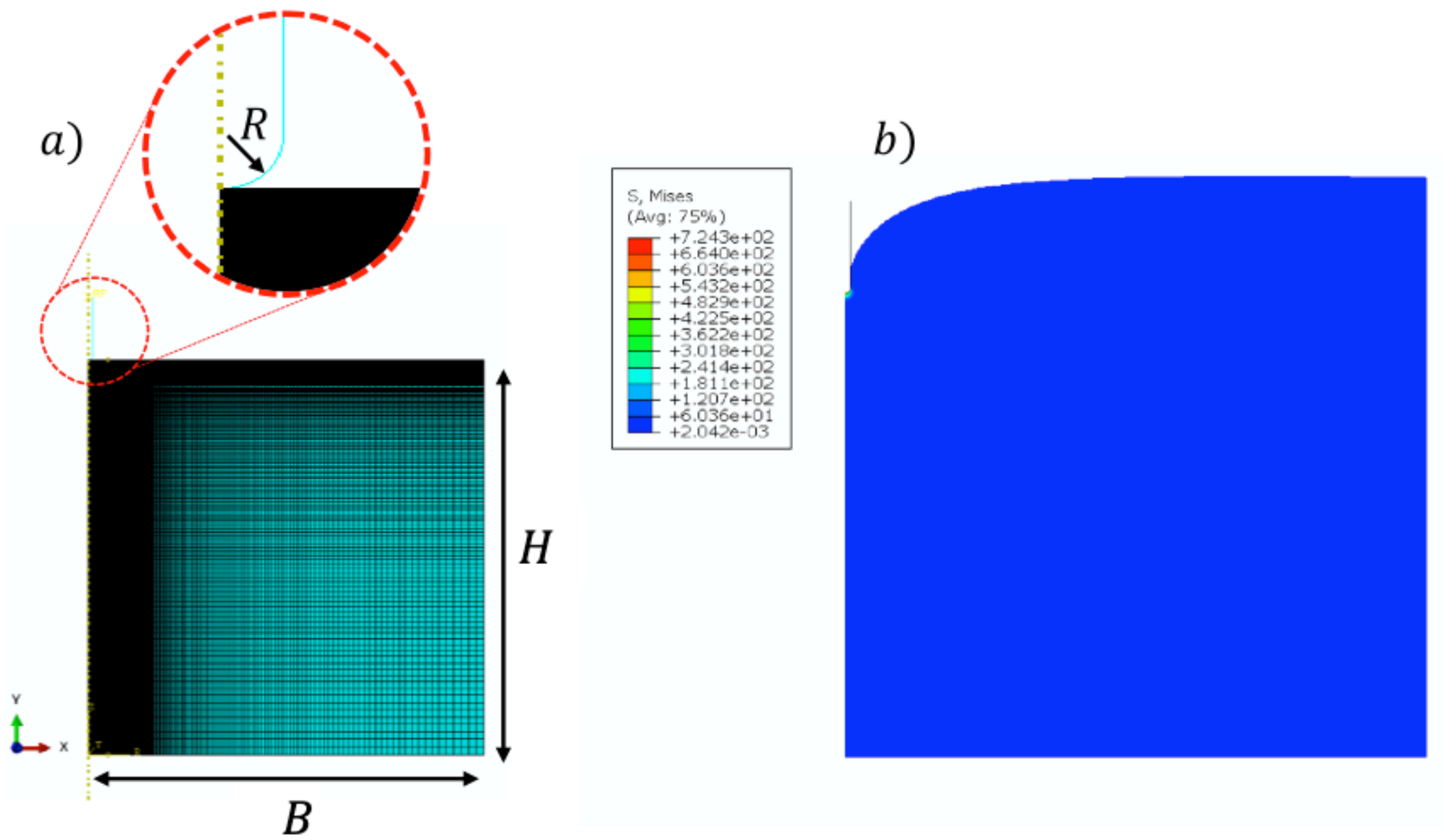

**Figure S1**: Sketch of the axisymmetric finite-element model for configuration, (a) showing the sample geometry and boundary conditions, with the initial mesh configuration, and (b) the deformed configuration under a single-term Ogden hyperelastic model ($\alpha = 9$) at $\mathrm{d/R} = 25$ indentation.

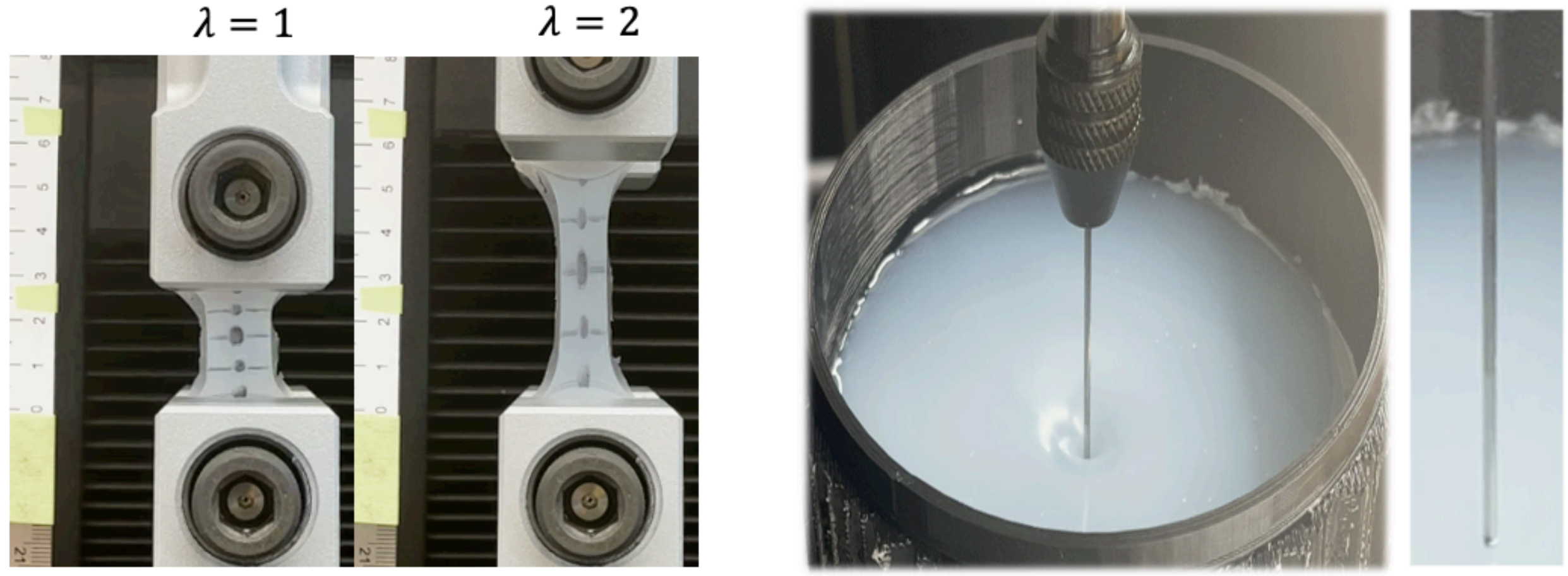

**Figure S2**: *Left:* Photographs of the dogbone-shaped specimen used in uniaxial tests, showing horizontal fiducial lines tracked optically to determine stretch. *Right:* Photographs of the spherical indenter (radius $R = 0.5\ mm$), composed of a steel ball glued to the end of a cylindrical steel shaft.